\documentclass[12pt,a4paper,epsf]{article}
\usepackage{graphics}
\usepackage{amssymb,amsmath}
\usepackage[dvips]{lscape,graphicx}
\usepackage{cite}
\usepackage{longtable}
\usepackage{slashed}
\usepackage{bbold}
\usepackage{soul}

\newcommand{\ct}{\cite}
\newcommand{\lb}{\label}

\newcommand{\bc}{\begin{center}}
\newcommand{\ec}{\end{center}}
\newcommand{\bd}{\begin{displaymath}}
\newcommand{\ed}{\end{displaymath}}
\newcommand{\be}{\begin{equation}}
\newcommand{\ee}{\end{equation}}
\newcommand{\ba}{\begin{array}}
\newcommand{\ea}{\end{array}}
\newcommand{\bea}{\begin{eqnarray}}
\newcommand{\eea}{\end{eqnarray}}
\newcommand{\bt}{\begin{tabular}}
\newcommand{\et}{\end{tabular}}

\newcommand{\bp}{\begin{picture}}
\newcommand{\ep}{\end{picture}}
\newcommand{\bfi}{\begin{figure}}
\newcommand{\efi}{\end{figure}}

\def\fun#1#2{\lower3.6pt\vbox{\baselineskip0pt\lineskip.9pt
\ialign{$\mathsurround=0pt#1\hfil##\hfil$\crcr#2\crcr\sim\crcr}}}

\mathchardef\mhyphen="2D 

\begin{document}
\vspace{0.5cm}

\title{\LARGE \bf {Degenerate Vacua of the Universe and What Comes
 Beyond the Standard Model}}
\author{\large\bf B.G.~Sidharth~$^1$\footnote{iiamisbgs@yahoo.co.in, birlasc@gmail.com} ,\; C.R.~Das~$^2$\footnote{das@theor.jinr.ru} ,\; C.D.~Froggatt~$^3$\footnote{Colin.Froggatt@glasgow.ac.uk} ,\\ \large\bf H.B.~Nielsen~$^4$\footnote{hbech@nbi.dk}\; and Larisa Laperashvili~$^5$\footnote{laper@itep.ru}\\\\
{\large \it $^1$~International Institute of Applicable Mathematics}\\
{\large \it and Information Sciences,}\\
{\large \it B.M. Birla Science Centre}\\
{\large \it Adarsh Nagar, 500063 Hyderabad, India}\\\\
{\large \it $^2$~Bogoliubov Laboratory of Theoretical Physics}\\
{\large \it Joint Institute for Nuclear Research}\\
{\large \it International Intergovernmental Organization,}\\
{\large \it Joliot-Curie 6, 141980 Dubna, Moscow region, Russia}\\\\
{\large \it $^3$~Glasgow University, UK}\\\\
{\large \it $^4$~Niels Bohr Institute,}\\
{\large \it Blegdamsvej, 17-21, DK 2100 Copenhagen, Denmark}\\\\
{\large \it $^5$~The Institute of Theoretical and
Experimental Physics,}\\
{\large\it National Research Center ``Kurchatov Institute'',}\\
{\large\it Bolshaya Cheremushkinskaya, 25, 117218 Moscow, Russia}}

\date{}
\maketitle

\thispagestyle{empty}

\newpage

\begin{abstract}

We present a new cosmological model of the Universe based on the
two discoveries: 1. cosmological constant is very small, and 2.
Nature shows a new law in physics called ``Multiple Point
Principle" (MPP). The MPP predicts the two degenerate vacua of the
Universe with VEV $v_1\approx 246$ Gev and $v_2\sim 10^{18}$ GeV,
which provide masses of the Higgs boson and top-quark. A new
cosmological model assumes the formation of two universal bubbles.
The Universe at first stage of its existing is a bubble with a de-Sitter
spacetime inside, having black-holes-hedgehogs as
topological defects of the vacuum. Such a bubble has a ``false
vacuum" with VEV $v_2$, which decays very quickly. Cooling
Universe has a new phase transition, transforming the ``false"
vacuum to the ``true" (Electroweak) vacuum. Hedgehogs confined, and
the universal bubble is transformed into the bubble having
spacetime with FLRW-metric and the vacuum with new topological
defects of $U(1)_{(el\mhyphen mag)}$ group: magnetic vortices and
Sidharth's pointlike defects. The problem of
stability/metastability of the EW-vacuum is investigated.
Noncommutativity of the vacua spacetime manifold is discussed. The
prediction of a new physics is given by the future observations at
LHC of the triplet $SU(2)$ Higgs bosons (at energies $E\sim 10$
TeV), and/or of the new bound states $6t + 6\bar t$ formed by
top-antitop quarks (at $E\sim 1$ TeV). The problem ``What comes
beyond the Standard Model" is discussed at the end of this paper.

\end{abstract}

\section*{Contents:}

\begin{enumerate}
{\bf
\item {\bf Dark Energy (DE) and Multiple Point Principle (MPP)}

a. {\bf Three vacua of the Standard Model (SM)}

b. {\bf Top-quark and Higgs boson mass prediction}

\item {\bf TOE, Graviweak Unification and Bubbles of the
Universe}

a. {\bf Big Bang}

b. {\bf Theory of Everything (TOE)}

c. {\bf Inflation of the Universe}

\item {\bf Black-hole-hedgehog's solutions}

a. {\bf Hedgehogs as topological defects of the ``false vacuum"}

\item {\bf Non-commutativity of the vacuum's spacetime manifold}

a. {\bf Sidharth's prediction for DE}

\item {\bf The phase transition from the ``false vacuum" to the
``true vacuum"}

\item {\bf Stability of the EW-vacuum}

\item {\bf The prediction of a New Physics}

\item {\bf Vacuum stability II}

\item {\bf What comes beyond the Standard Model}
}
\end{enumerate}

\newpage

\section*{Introduction}

In this paper we present a new standard cosmological model which
is based on the two discovery of the XX(end)-XXI(beginning)
century:\begin{enumerate} \item cosmological constant (Dark Energy) is extremely
small, and
\item Nature shows a ``new law of physics" which is named Multiple
Point Principle (MPP).\\ \\ {\ul {Multiple Point Principle}}
postulates:\\ \\ {\it There are several vacua in Nature with the same
energy density, or cosmological constant, and all cosmological
constants are zero, or approximately zero.}\end{enumerate}

\section{Dark Energy (DE) and Multiple Point Principle (MPP)}

Multiple Point Principle (MPP) was first suggested by D.L. Bennett
and H.B. Nielsen in Ref.~\ct{1}.

A priori it is quite possible for a quantum field theory to have
several minima of its effective potential as a function of its
scalar fields (in our paper -- scalar Higgs bosons). Postulating
zero cosmological constant, we confront ourselves with a question:
should the energy density, i.e. the cosmological constant, be (at
least approximately) zero for all possible vacua or should it only
be zero for that vacuum in which we live? The assumption would not
be more complicated, if we postulated that all the vacua which
might exist, as minima of the effective potential, should have
approximately zero cosmological constant.

The MPP theory was developed in a lot of papers by H.B. Nielsen,
D.L. Bennett, C.D. Froggatt, R.B. Nevzorov, L.V. Laperashvili,
C.R. Das, and recently in Refs.~\ct{2,3,4}.

Vacuum energy density of our Universe is the Dark Energy (DE),
which is related with cosmological constant $\Lambda$ by the
following way:
\be \rho_{DE} = \rho_{vac} = (M^{red}_{Pl})^2\Lambda. \lb{1} \ee
Here $M^{red}_{Pl}$ is the reduced Planck mass: $M^{red}_{Pl}\simeq 2.43 \times 10^{18}$ GeV.

Recent cosmological measurements (see Particle Data Group
\ct{PDG}) give:
\be \rho_{DE} \simeq (2\times 10^{-3}\; {\rm{ eV}})^4. \lb{2} \ee
According to (\ref{2}), we have a very small value of cosmological
constant $\Lambda$:
\be \Lambda \simeq 10^{-84}\; {\rm{ GeV}}^2. \lb{3} \ee
This tiny value of $\rho_{DE}$ was first predicted by B.G.
Sidharth in 1997 \ct{5,6}, explaining that the Universe has an
accelerating expansion. In 2011 S. Perlmutter, B. Schmidt and A.
Riess were awarded by the Nobel Prize for discovery of the
Universe accelerating expansion.

Considering extremely small cosmological constant of
our Universe, Bennett, Froggatt and Nielsen \ct{1,10,11} assumed
only zero, or almost zero, cosmological constants for all vacua
existing in the Universe.

\subsection*{a. Three vacua of the Standard Model (SM)}

Restricted ourselves to {\bf the pure Standard Model (SM)} we have only three vacua:\begin{enumerate}
\item {\bf Present Electroweak vacuum}, in which we live.\\
It has vacuum expectation value (VEV) of the Higgs field equal to:
\be v_1 = v = \langle\phi_H\rangle \approx 246 \;{\rm{GeV}}. \lb{4} \ee
\item {\bf High Higgs field vacuum} -- Planck scale vacuum, which has
the following VEV:
\be v_2 = v = \langle\phi_H\rangle \sim 10^{18} \;{\rm{GeV}}. \lb{5} \ee
\item {\bf Condensate vacuum.} This third vacuum is a very
speculative possible state inside the pure SM, which contains a
lot of strongly bound states, each bound from 6 top + 6 anti-top
quarks (see Refs.~\ct{7,8,9}).\end{enumerate}

From experimental results for these three vacua, cosmological
constants which corresponds to the minimum of the Higgs effective
potential $V_{eff} (\phi_H)$, are not exactly equal to zero.
Nevertheless, they are extremely small. By this reason, Bennett,
Froggatt and Nielsen \ct{1,10,11} assumed to consider zero
cosmological constants as a good approximation. Then according to
the MPP, we have a model of the pure SM being finetuned in such a
way that these three vacua proposed have
just zero energy density.

If the effective potential has three degenerate minima, then the
following requirements are satisfied \ct{10,11}:
\be V_{eff} (\phi^2_{min1}) = V_{eff} (\phi^2_{min2}) = V_{eff}
(\phi^2_{min3}) = 0, \lb{6} \ee
and \be V'_{eff} (\phi^2_{min1}) = V'_{eff} (\phi^2_{min2}) =
V'_{eff} (\phi^2_{min3}) = 0, \lb{7} \ee
where
\be V'(\phi^2) = \frac{\partial V}{\partial \phi^2}. \lb{8} \ee
Here we assume that:
$$ V_{eff} (\phi^2_{min1}) = V_{present},$$
$$ V_{eff} (\phi^2_{min2}) = V_{high\;field},$$
and $$ V_{eff} (\phi^2_{min3}) = V_{condesate}. $$

\subsection*{b. Top-quark and Higgs boson mass prediction.}

Assuming the existence of the two degenerate vacua in the SM:
\begin{enumerate}
\item the first Electroweak vacuum at $v_1 \approx 246$ GeV, and
\item the second Planck scale vacuum at $v_2 \sim 10^{18}$ GeV,\end{enumerate}
Froggatt and Nielsen predicted the top-quark and Higgs boson
masses \ct{10}:
\be M_t = 173 \pm 5\; {\rm { GeV}}; \quad M_H = 135 \pm 10\;
{\rm { GeV}}. \lb{10} \ee
In Fig.~1 it is shown the existence of
the second (non-standard) minimum of the effective Higgs potential
in the pure SM at the Planck scale.

\section{TOE, Graviweak Unification and Bubbles of the
Universe}

\subsection*{a. Big Bang}

With aim to explain why we have two universal vacua, let us start
with the Big Bang.

The Universe was expanded from a very high density and high
temperature state. Big Bang is a singularity, the result of the
extrapolation of all known laws of physics to the highest density
regime. This primordial singularity called ``the Big Bang" is the
``birth" of our Universe since it represents the point in time when
the Universe entered into a regime where the laws of physics
(General Relativity, SM, etc.) began to work.

The time that has passed since that event is known as {\ul{``the
age of the universe"}} $T_U$: $$T_U\simeq 13.799 \pm 0.021\;
{\rm{billion\; years}}.$$

\subsection*{b. Theory of Everything (TOE)}

Through years of research scientists learned that GR and QFT are
mutually incompatible: they cannot both be right. This
incompatibility between GR and QFT is essential only in regions of
extremely small-scale and high-mass, which exist during the
beginning stages of the Universe, from the moment immediately
following the Big Bang.

To resolve this problem, theorists tried to construct a theory
unifying gravity with the other three interactions. It was
searching a single theory that is capable of describing all
phenomena. This is a {\ul{``Theory of Everything" - TOE}}.

This term was used by John Ellis in his article published in
``Nature" in 1986.

In this goal, quantum gravity has become an area of successful
research. The Superstring Theory intends to be the ultimate theory
of the Universe - TOE.

On November 6, 2007, Antony Garrett Lisi (Hawai Univ.) \ct{12}
suggested ``An Exceptionally Simple Theory of Everything", often
referred to as ``$E_8$ Theory", which attempts to describe all known
fundamental interactions in physics using the Lie algebra of the
largest ``simple", ``exceptional" Lie group, $E_8$. The paper
describes how the combined structure and dynamics of all
gravitational and Standard Model particle fields, including
fermions, are part of the $E_8$ Lie algebra. See elementary
particle states assigned to $E_8$ roots in Fig.~2.

The $E_8$ Lie group has applications in theoretical physics and
especially in String theory and Supergravity.

$E_8\times E_8$ is the gauge group of one of the two types of
heterotic string. Heterotic string theory was developed in 1985
(see Ref.~\ct{13}) by David Gross, Jeffrey Harvey, Emil Martinec,
and Ryan Rohm (the so-called ``Princeton String Quartet").

\subsection*{c. Inflation of the Universe}

TOE contains a statement that at the beginning of the Universe, up
to $10^{-43}$ seconds after the Big Bang, the four fundamental
forces were once a single fundamental force.

Approximately in $10^{-37}$ seconds, a phase transition caused a
cosmic inflation, and the Universe began to grow exponentially
during which time density fluctuations (occurred because of the
uncertainty principle) were amplified into the seeds that would
later form the large-scale structure of the Universe.

In Refs.~\ct{14,15} we have suggested the {\ul{Graviweak
unification}} with a group of symmetry $$G_{(GW)} = Spin(4,4),$$
which is spontaneously broken into the
$$SL(2,C)^{(grav)}\times SU(2)^{(weak)}.$$
We assumed that after the Big Bang there existed a Theory of
Everything (TOE) which rapidly was broken down to the direct
product of the following gauge groups:
\bea G_{(TOE)} &\to& G_{(GW)}\times U(4) \to SL(2,C)^{(grav)}\times SU(2)^{(weak)}\times U(4)\nonumber\\
&\to& SL(2,C)^{(grav)}\times SU(2)^{(weak)}\times SU(4)\times U(1)_Y\nonumber\\
&\to& SL(2,C)^{(grav)}\times SU(2)^{(weak)}\times SU(3)_c\times U(1)_{(B-L)}\times U(1)_Y\nonumber\\
&\to& SL(2,C)^{(grav)}\times SU(3)_c\times SU(2)_L \times U(1)_Y\times U(1)_{(B-L)}\nonumber\\
&\to& SL(2,C)^{(grav)}\times G_{SM} \times
U(1)_{(B-L)}.\nonumber\eea And below the see-saw scale ($M_R\sim
10^9\to 10^{14}$ GeV) we have the SM group of symmetry:
$$ G_{SM} = SU(3)_c\times SU(2)_L \times U(1)_Y.$$

\section{Black-hole-hedgehog's solutions}

The action $S_{(GW)}$ of the Graviweak unification (obtained in
our papers) is given by the following expression:
\bea S_{(GW)} &=& - \frac 1{g_{uni}}\int_{\mathfrak
M}d^4x\sqrt{-g}\left[\frac 1{16}\left(R|\Phi|^2 - \frac 32|\Phi|^4\right)\right.\nonumber\\
&&+\left. \frac 1{16}\left(aR_{\mu\nu}R^{\mu\nu} + bR^2\right) +
\frac 12{\cal D}_{\mu}\Phi^\dag{\cal D}^{\mu}\Phi + \frac 14
F_{\mu\nu}^iF^{i\,\mu\nu}\right], \lb{15} \eea
where $g_{uni}$ is a parameter of the graviweak unification,
parameters $a,b$ (with $a+b=1$) are ``bare" coupling constants of
the higher derivative gravity, $R$ is the Riemann curvature
scalar, $R_{\mu\nu}$ is the Ricci tensor, $|\Phi|^2 =
\Phi^a\Phi^a$ is a squared triplet Higgs field, where $\Phi^a$
(with $a=1,2,3$) is an isovector scalar belonging to the adjoint
representation of the $SU(2)$ gauge group of symmetry. In
Eq.~(\ref{15}):
\be {\cal D}_{\mu}\Phi^a = \partial_{\mu}\Phi^a +
g_2\epsilon^{abc}A_{\mu}^b\Phi^c \lb{16} \ee
is a covariant derivative, and
\be F_{\mu\nu}^a = \partial_{\mu}A_{\nu}^a -
\partial_{\nu}A_{\mu}^a + g_2\epsilon^{abc}A_{\mu}^bA_{\nu}^c \lb{17} \ee
is a curvature of the gauge field $A_{\mu}^a$ of the $SU(2)$
Yang-Mills theory. The coupling constant $g_2$ is a ``bare"
coupling constant of the $SU(2)$ weak interaction.

The GW action (\ref{15}) is a special case of {\ul{the $f(R)$-gravity}} when:
\be f(R) = R|\Phi|^2. \lb{18} \ee
In a general case of the $f(R)$-gravity, the action can be
presented by the following expression:
\be S = \frac{1}{2\kappa}\int d^4x \sqrt{-g}\,f(R) + S_{grav} +
S_{gauge} + S_m, \lb{19} \ee
where $S_m$ corresponds to the part of the action associated with
matter fields (fermions and Higgs fields).

Using the metric formalism, we obtain the following field
equations:
\be F(R)R_{\mu\nu} - \frac 12 f(R)g_{\mu\nu} - \nabla_{\mu}
\nabla_{\nu}F(R) + g_{\mu\nu}\Box F(R) = \kappa T_{\mu\nu}^m,
 \lb{20} \ee
where:
\be F(R)\equiv \frac{df(r)}{dr}, \lb{21} \ee
$\kappa = 8\pi G_N$, $G_N$ is the gravitational constant, and
$T^m$ is the energy-momentum tensor derived from the matter action
$S_m$.

Here we must emphasize a very important point of the present
theory:\\ \\ the ``$E_8$ Theory" (TOE), and consequently the Graviweak
unification, contain the {\ul{triplet $SU(2)$ Higgs field
$\Phi^a$}}, in contrast to the SM, which considers {\ul{doublet
$SU(2)$ Higgs field $H^{\alpha}$}}.

Just this field $\Phi^a$ is responsible for the formation of the
gravitational black-holes-hedgehogs inside a Bubble of the
Universe with a de-Sitter spacetime inside, having hedgehogs as
vacuum topological defects. The field $\Phi^a$ constructs the
black-hole-hedgehog's solutions in the GWU $f(R)$-gravity.

Following to the idea by A.Vilenkin \ct{16}, it is possible to
consider the Universe at first stage of its existing as a bubble
with a de-Sitter spacetime inside, having global monopoles as
vacuum topological defects.

Hedgehogs are extended objects, which are global monopoles. They
have repulsive forces of interaction, which lead to the inflation
of the universal bubble. Such a bubble has a vacuum with a Planck
scale VEV:
$$v_2 \sim 10^{18} \;{\rm{GeV}}.$$
This vacuum decays very quickly, and by this reason is called the
{\ul{``false vacuum"}}.

A global monopole is described by the part $L_h$ of the
Lagrangian $L_{(GW)}$, which contains the $SU(2)$-triplet Higgs
field $\Phi^a$, VEV of the second vacuum $v_2=v$ and cosmological
constant $\Lambda=\Lambda_E$:
$$ L_h = - \frac{R}{16}|\Phi|^2 + \frac{3g_2^2}{32}|\Phi|^4 - \frac
12\partial_{\mu}\Phi^a\partial^{\mu}\Phi^a + \Lambda_E $$ \be = -
\frac 12
\partial_{\mu}\Phi^a\partial^{\mu}\Phi^a +
\frac{\lambda}{4}\left(|\Phi|^2 - v^2\right)^2 +
\frac{\Lambda_E}{\kappa} - \frac{\lambda}{4}v^4 = - \frac 12
\partial_{\mu}\Phi^a\partial^{\mu}\Phi^a +
\frac{\lambda}{4}\left(|\Phi|^2 - v^2\right)^2. \lb{41} \ee
Here we have: $ \lambda = \frac {3g_2^2}{8}.$

The field configurations describing a monopole-hedgehog are:
\be \Phi^a = v w(r )\frac{x^a}{r}, \lb{1m} \ee
\be A_{\mu}^a = a(r )\epsilon_{\mu ab}\frac{x^b}{r}, \lb{2m} \ee
where $x^ax^a = r^2$ with $(a = 1, 2, 3)$, $w(r)$ and $a(r)$ are
some structural functions. This solution is pointing radially.
Here $\Phi^a$ is parallel to $\hat{r}$ -- the unit vector in the
radial, and we have a ``hedgehog" solution of Refs.~\ct{17,18}.
The terminology ``hedgehog" was first suggested by Alexander
Polyakov \ct{18}.

In Ref.~\ct{19} by solving the gravitational field equations we
estimated the black-hole-hedgehog's mass $M_{BH}$, radius $R_{BH}$
and horizon radius $r_h$:

$M_{BH}\approx 3.65\times 10^{18}$ GeV, \quad $R_{BH} \sim 10^{-21}$ GeV$^{-1}$,
\quad and\quad
$r_h \approx 2.29R_h$ respectively.

We obtained a ``hedgehog" -- global monopole, that has been
``swallowed" by the black-hole with mass core $M_{BH}\sim 10^{18}$
GeV and radius $\delta\sim 10^{-21}$ GeV$^{-1}$.

\subsection*{a. Hedgehogs as topological defects of the ``false
vacuum" }

We see that at the first stage of the evolution, the Universe is a
Bubble with a de-Sitter spacetime inside, and the Universe radius
is close to the de-Sitter horizon radius:

 $R_{UN}\simeq R_{de\mhyphen Sitter\; horizon} \simeq 10^{28}$ cm.

The vacuum of this Bubble which is a ``false vacuum" of the
Universe, contains black-hole-hedgehogs as topological defects.

Now the vacuum reminds a boiling water with little bubbles of
vapor.

A global monopole is a heavy object formed as a result of the
gauge-symmetry breaking during the phase transition of the
isoscalar triplet $\Phi^a$ system. The black-holes-hedgehogs are
similar to elementary particles, because a major part of their
energy is concentrated in a small region near the monopole core.

\section{Non-commutativity of the vacuum's spacetime manifold}

Assuming that the Planck scale false vacuum is described by a
non-differentiable space-time having lattice-like structure,
where sites of the lattice are black-holes with ``hedgehog"
monopoles inside them, we describe this manifold by a
non-commutative geometry with a minimal length $l=\lambda_{Pl}$.

In the non-commutative geometry coordinates obey the following
commutation relations:
\be [dx^{\mu}, dx^{\nu}] \approx \beta^{\mu\nu}l^2 \neq 0,
\lb{20a} \ee
containing any minimal cut off $l$.

Previously the following commutation relation was considered by
H.S. Snyder \ct{20}
\be [x, p] = \hslash \left( 1 + \left(\frac{l}{\hslash}
\right)^2p^2\right),\; etc., \lb{21a} \ee
which shows that effectively 4-momentum $p$ is replaced by
\be p \to p\left( 1 +
\left(\frac{l}{\hslash}\right)^2p^2\right)^{-1}.
 \lb{22a} \ee
Applied to the Compton wavelength this gives the so called
{\ul{Snyder-Sidharth dispersion relation}}:
\be [x_\imath , x_j] = \beta_{\imath j} \cdot l^2, \lb{23a} \ee
which leads to a modification in the Dirac and also in the
Klein-Gordon equation, as described in details in Ref.~\ct{21}.

The modified Dirac equation is
\be \left\{\gamma^\circ p^\circ + \Gamma + \gamma^5 \alpha l
p^2\right\} \psi = 0, \lb{24a} \ee
which contains an extra term. The extra term gives a slight mass
for the neutrino which is roughly of order $\sim 10^{-8} m_e$,
where $m_e$ being the mass of the electron. Thus, the
non-commutative geometry gives a Lagrangian describing the
electron neutrino mass $m_{\nu_e}$.

\subsection*{a. Sidharth's prediction for DE}

Using the non-commutative theory of the discrete space-time, B.G.
Sidharth predicted in Ref.~\ct{6} (see also the book \ct{22}) a
tiny value of cosmological constant:
 $$\Lambda \simeq 10^{-84}\;{\rm{GeV^2}}$$
as a result of the compensation of Zero Point Fields contributions
by non-commutative contributions of the vacuum lattice.

\section{The phase transition from the ``false vacuum" to the ``true
vacuum"}

At the early stage, the Universe is very hot, but then it begins
to cool down. Black-holes-monopoles (as bubbles of the vapor in
the boiling water) begin to disappear. The temperature dependent
part of the energy density dies away. In that case, only the
vacuum energy density can survive. Since this is a constant, the
Universe expands exponentially, and an exponentially expanding
Universe leads to the inflation. While the Universe is expanding
exponentially, so it is cooling exponentially. This scenario was
called {\ul{supercooling}} of the Universe. When the temperature
reached the critical value $T_c$, the Higgs mechanism of the SM
created a new condensate $\phi_{min1}$, and the vacuum became
similar to superconductor, in which the topological defects are
magnetic vortices.

The energy of black-holes is released as particles, and all these
particles (quarks, leptons, vector bosons) acquired their masses
$m_i$ through the Yukawa coupling mechanism $Y_f \bar
\psi_f\psi_f\phi$. Therefore, they acquired the Compton
wavelength, $\lambda_i=\hbar/m_ic$. And according to the
Sidharth's theory, in the EW-vacuum we again have lattice-like
structures formed by bosons and fermions, and the lattice
parameters ``$l_i$" are equal to the Compton wavelengths: $l_i =
\lambda_i = \hbar/m_ic$.

At some finite cosmic temperature which is the critical
temperature $T_c\sim 10^{15}\;K^0$, a system exhibits a
spontaneous symmetry breaking, and we observe a phase transition
from the bubble with the false vacuum to the bubble with the true
vacuum.

Hedgehogs confined, and universal Bubble is transformed into the
bubble having a spacetime with FLRW-metric
(Friedmann-Lematre-Robertson-Walker metric), and vacuum acquires
new topological defects. These new topological defects belong to
the $U(1)_{(el\mhyphen mag)}$ group. They are:\begin{enumerate}\item magnetic
vortices--``strings" by Abrikosov-Nielsen-Olesen \ct{23,24}, and\item
b. Sidharth's Compton wave topological objects \ct{25,26}.\end{enumerate}

After the phase transition, the Universe begins its evolution
toward the low energy Electroweak (EW) phase. Here the Universe
undergoes the inflation, which leads to the phase having the VEV:
$$v_1\approx 246\;{\rm{GeV}}.$$
This is a ``true" vacuum, in which we live.

\section{Stability of the EW-vacuum}

Here we must emphasize that due to the energy conservation law,
the vacuum energy density before the phase transition (for $T >
T_c$) is equal to the vacuum energy density after the phase
transition (for $T < T_c$), therefore we have:
\be \rho_{vac}({\rm at\; Planck\; scale}) = \rho_{vac}({\rm
at\; EW\; scale}). \lb{12pt} \ee
The analogous link between the Planck scale phase and EW phase was
considered in Ref.~\ct{27}. It was shown that the vacuum energy
density (DE) is described by the different contributions to the
Planck and EW scale phases. This difference is a result of the
phase transition. However, the vacuum energy densities (DE) of
both vacua are equal, and we have a link between gravitation and
electromagnetism via the Dark Energy. Here we see: since
$\rho_{vac}$ (at the Planck scale) is almost zero, then
$\rho_{vac}$ (at EW scale) also is almost zero, and we have a
triumph of the Multiple Point Principle: we have the two
degenerate vacua with almost zero vacuum energy density.

\section{The prediction of a New Physics}

In our model we investigated hedgehogs in the Wilson loops of the
$SU(2)$ Yang-Mills theory using the results of Ref.~\ct{28}.
Considering their lattice result for the critical value of the
temperature of hedgehog's confinement phase: $$\beta_{crit}
\approx 2.5$$ (here $\beta=1/g_2^2$) we predicted in our recent
paper (see Ref.~\ct{19}) the production of the $SU(2)$-triplet
Higgs bosons $\Phi^a$ at LHC at energy scale $$\mu \sim
10\;{\rm{TeV}}.$$ At this energy we can expect to see at LHC the
production of the triplet Higgs bosons with mass $\gtrsim 5$ TeV.

This provides a new physics in the SM.

\section{Vacuum stability II}

Taking into account that hedgehog fields $\Phi^a$ produce a new
physics at the scale $\sim 10$ TeV, we considered an additional
confirmation of the vacuum stability and accuracy of the MPP.

As it was mentioned at the beginning of this paper, Froggatt and
Nielsen predicted the top-quark and Higgs boson masses: $M_t = 173
\pm 5$ GeV and $M_H = 135 \pm 10$ GeV, assuming the existence of
two degenerate vacua in the SM (the first Electroweak vacuum and
the second Planck scale one).

Their prediction for the mass of the SM $SU(2)$-doublet Higgs
boson was improved in Refs.~\ct{29,30}. They gave calculations of
the 2-loop and 3-loop radiative corrections to the effective Higgs
potential $V_{eff}(H)$. The prediction of G. Degrassi at al. is:
$M_H = 129 \pm 2$ GeV, which provides the possibility of the
theoretical explanation of the value $M_H \simeq 125.7$ GeV
observed at LHC.

From Degrassi et al. calculation, the effective Higgs field
potential $V_{eff}(H)$ has a minimum, which slightly goes under
zero, so that the present EW-vacuum is unstable for the
experimental Higgs mass $M_H \simeq 125.7 \pm 0.24$ GeV.

The position of the second minimum depends on the SM parameters,
especially on the top and Higgs masses, $M_t$ and $M_H$. This
$V_{eff}(min2)$ can be higher or lower than the $V_{eff}(min1)$
showing a stable EW vacuum (in the first case), or metastable one
(in the second case).

The red solid line of Fig.~3 by Degrassi et al. \ct{29} shows the
running of the Higgs self-interaction coupling constant
$\lambda_{H,eff}(\phi)$ for $M_H \simeq 125.7$ GeV and $M_t \simeq
171.43$ GeV, which just corresponds to the Borderline vacuum
stability, that is, to the stable EW-vacuum. In this case the
minimum of the $V_{eff}(H)$ exists at the $\phi = \phi_0 \sim
10^{18}$ GeV, where according to MPP:
$$ \lambda_{H,eff}(\phi_0) = \beta(\lambda_{H,eff}(\phi_0)) = 0.$$
Unfortunately, this case does not correspond to the current
experimental values.

In Fig.~3 blue lines (thick and dashed) present the RG evolution
of $\lambda_H(\mu)$ for current experimental values $M_H \simeq
125.7$ GeV and $M_t\simeq 173.34$ GeV. The thick blue line
corresponds to the central value of $\alpha_s = 0.1184$ and dashed
blue lines correspond to its errors equal to $\pm 0.0007$. We see
that absolute stability of the Higgs potential is excluded by at
98\% C.L. for $M_H < 126$ GeV.

Fig.~3 shows that asymptotically $\lambda_H(\mu)$ does not reach
zero but approaches to the negative value:
\be \lambda_H \to - 0.01\pm 0.002, \lb{31n} \ee
indicating the metastability of the EW vacuum. We see that the
current experimental values of $M_H$ and $M_t$ show the
metastability of the present EW-vacuum of the Universe, and this
result means that the MPP law is not exact.

Can the MPP be exact due to the corrections from hedgehogs'
contributions? We think that it is possible.

If we assume that in the region $E > E_{threshold}$ the effective
Higgs potential contains not only the $SU(2)$-triplet field
$\Phi^a$, but also the $SU(2)$-doublet Higgs field $H^{\alpha}$
(where $a = 1,2,3$ and $\alpha = 1,2$), then there exists an
interaction (mixing term) between these two Higgs fields (see
\ct{19}). Of course, the effective Higgs self-interaction coupling
constant $\lambda_{H,\;eff}(\mu)$ is a running function
presenting loop corrections to the Higgs mass $M_H$, which arise
from the Higgs bosons $H$ $(\Delta\lambda_H(\mu))$ and from
hedgehogs $h$ $(\delta\lambda_H(\mu))$:
\be \lambda_{H,eff}(\mu) = \frac{G_F}{\sqrt 2}M_H^2 +
\Delta\lambda_H(\mu) + \delta\lambda_H(\mu), \lb{33n} \ee
where $G_F$ is the Fermi constant. The main contribution to the
correction $\delta\lambda_H(\mu)$, described by a series in the
mixing coupling constant $\lambda_{hH}$, is a term $\lambda_S$
given by the Feynman diagram of Fig.~4 containing the hedgehog $h$
in the loop:
\be \delta\lambda_H(\mu) = \lambda_S(\mu) + .... \lb{34n} \ee
Here the effective Higgs self-interaction coupling constant
$\lambda_{H,eff}(\mu)$ is equal to $\lambda_{eff}(\mu)$ considered
by Degrassi et al.

Our hedgehog is an extended object with a mass $M_h$ and radius
$R_h$, therefore it is easy to estimate $\lambda_S$ at high
energies $\mu > E_{threshold}$ by methods described in our paper
\ct{31}. And we obtained:
\be \lambda_S(\mu) \approx \frac
1{16\pi^2}\frac{\lambda_{hH}^2(\mu)}{(R_h M_h)^4}, \lb{35n} \ee
where $\lambda_{hH}(\mu)$ is a running coupling constant of the
interaction of hedgehogs $h$ with the Higgs fields $H$. In
Eq.~(\ref{35n}) parameters $M_h=M_{BH}$ and $R_h=R_{BH}$ are the
running mass and radius of the hedgehog, respectively.

As we have shown in \ct{19}, at high Planck scale energies they
are:
\be M_h \sim 10^{18}\; {\rm{GeV}},\quad
 R_h\sim 10^{-21}\; {\rm{GeV^{-1}}}, \lb{37n} \ee
and
\be R_h M_h \sim 10^{-3}. \lb{38n} \ee
As a result, we have:
\be \lambda_S \sim \frac{{\lambda_{hH}}^2}{16\pi^2}10^{12}.
\lb{39n} \ee
If hedgehog parameter $\lambda_{hH}$ is:
\be \lambda_{hH}\sim 10^{-6}, \lb{40n} \ee
then
\be \lambda_S\sim 0.01, \lb{41n} \ee
and the hedgehogs' contribution transforms the metastable (blue)
curve of Fig.~3 into the ``Borderline vacuum stability MPP" (red)
curve, and we have an exact stability of the EW-vacuum and the
exact MPP, that is, two degenerate vacua in the Universe.

\section{What comes beyond the Standard Model}

Standard Model of particle physics is the most complete theory. In
this paper we present our (non trivial) efforts to go beyond the
Standard Model (SM). We try to overcome the following shortcomings
of the SM (see also Ref.~\ct{32}):

\begin{enumerate}
\item SM doesn't include gravity, the fourth fundamental interaction.

\item SM doesn't deliver the mass of neutrino:\\ neutrino remains a
massless particle in the SM.

\item SM can be changed by the existence of new yet undiscovered
particles:

\begin{enumerate}
\item by Supersymmetry which presents the supersymmetric counterparts
of the SM particles;

\item by the existence of more heavy multiplets of the SM group
$G_{(SM)}$;

\item by the existence of new bound states (NBS) in the framework of
the SM, for example, by the existence $6t+6\bar t$ NBS, suggested
in Refs.~\ct{7,8,9}.
\end{enumerate}

\item There is no place for Dark Energy in the SM.

\item SM doesn't describe Dark Matter.

\item It is also difficult in the SM to accommodate the observed
predominance of matter over antimatter (matter/antimatter
asymmetry).

\item Finally, the SM cannot explain the 19 arbitrary constants which
are contained in theory.
\end{enumerate}
Earlier, the problem of the SM near the Planck scale was
considered in the review \ct{33}.

Going beyond the SM we are able to explain some points of the SM
shortcomings:

\begin{enumerate}
\item In the present theory gravity is included by consideration of
the Graviweak Unification model \ct{14,15}.

\item The mass of neutrino is given by a theory of non-commutativity,
applied to the universal vacuum by B.G. Sidharth \ct{21,34}, who
considered the modified Dirac equation and predicted the mass of
neutrino:
$$ m_{\nu}\sim 10^{-8}m_e,$$
where $m_e$ is an electron mass.
\item
\begin{enumerate}  
\item The present theory predicts that Supersymmetry cannot be
observed at LHC because of the very high scale of SUSY:
$M_{SUSY}\sim 10^{18}$ GeV.

\item In Ref.~\ct{19}, reviewed in this paper, we predicted the
production of the $SU(2)$ triplet Higgs bosons at energy $\sim 10$
TeV, which can be detected by LHC.

\item We also suggested a theory, which predicts the existence of
the new bound states (NBS) created by the interaction of the SM
Higgs bosons with 6 top and 6 antitop quarks \ct{7,8,9}. Such
$6t+6\bar t$ resonances can be observed by LHC at energy $\sim 1$
TeV \ct{35}.
\end{enumerate}
\item Sidharth's theory of non-commutativity applied to the universal
vacuum spacetime manifold, gives an explanation of the DE
\ct{5,6}. And although Supergravity also explains the smallness of
the cosmological constant (i.e. DE) (see Ref.~\ct{36}), the
present theory suggests the Sidharth's explanation of the origin
of DE.

\item There are a lot of different theories published in the world
literature which are devoted to the origin of Dark Matter, but it
has not yet been definitely found.

A very interesting possibility is to consider the DM as a matter
of the Hidden World (HW), where HW is a Mirror World with broken
mirror parity (see for example Refs.~\ct{37,38,39,40}).

\item The Hidden World can explain the observed predominance of
matter over antimatter (matter/antimatter asymmetry) \ct{42,43}.

But here we also have difficulties, and Dark matter still
continues to be a mysterious phenomenon in cosmology.

\item Finally, 19 parameters of the SM can be described by Multiple
Point Model (see attempts in Refs.~\ct{7,41}).

\item And Hierarchy problem can be solved (see Refs.~\ct{44,45}).
\end{enumerate}

In conclusion, we wish to emphasize that the present cosmological
model shows new possibilities in cosmology.

\section{Conclusions}

In this paper:

\begin{enumerate}

\item We have shown that the evolution of the Universe from the
Grand Unification $E_8$ (TOE) gives a possibility to construct a
quite new cosmological model.

\item This new cosmological model is based on the two discovery:

1. cosmological constant (Dark Energy) is extremely small, and

2. Nature shows a ``new law of physics" which is named Multiple
Point Principle (MPP). MPP was first suggested by D.L. Bennett and
H.B. Nielsen, and postulates: {\it There are several vacua in
Nature with the same energy density, or cosmological constant, and
all cosmological constants are zero, or approximately zero.}

\item We have considered that vacuum energy density $\rho_{vac}$
of our Universe is the Dark Energy (DE), which is related with
cosmological constant $\Lambda$: $\rho_{DE} = \rho_{vac} =
(M^{red}_{Pl})^2\Lambda$, that a tiny value of $\rho_{DE}$ was
first predicted by B.G. Sidharth in 1997 who showed that
$\Lambda\simeq 10^{-84}\; {\rm GeV}^2$. B.G. Sidharth predicted:
that this very small cosmological constant (and DE-density)
provides an accelerating expansion of our Universe after the Big
Bang.

\item We confirmed the existence of the two degenerate vacua in
the SM: a. the first Electroweak vacuum at $v_1 = 246$ GeV, which
is a ``true" vacuum, and b. the second ``false" vacuum at the
Planck scale with VEV $v_2 \sim 10^{18}$ GeV.

\item We have shown that assuming the existence of the two
degenerate vacua in the SM, Froggatt and Nielsen predicted the
top-quark and Higgs boson masses: $M_t = 173 \pm 5$ GeV and $M_H =
135 \pm 10$ GeV. Their prediction for the top quark mass $M_t$ was
confirmed by SLAC with great accuracy. But the LHC result for the
discovered Higgs boson: $M_H \approx 125.7$ came in 2012.

\item The Froggatt-Nielsen prediction of the mass of the SM
$SU(2)$-doublet Higgs boson was improved by calculations of the
2-loop and 3-loop radiative corrections to the effective Higgs
potential $V_{eff}(H)$. Their prediction: $M_H = 129 \pm 2$ GeV
provided the theoretical explanation of the value $M_H \simeq
125.7$ GeV observed at LHC.

\item We described the evolution of Bubbles of the Universe. We
have shown that it is possible to consider the Universe at first
stage of its existing as a bubble with a de-Sitter spacetime
inside, having hedgehogs as vacuum topological defects. The
space-time inside the bubble with ``the true vacuum", has the
geometry of an open FLRW universe. The low-energy ``true vacuum" is
Electroweak (EW) vacuum, in which we live. This vacuum has
topological defects of $U(1)_{(el\mhyphen mag)}$ group: the
Abrikosov-Nielsen-Olesen magnetic vortices (``ANO strings") and
Sidharth's Compton wave objects.

\item We demonstrated that after the Big Bang Universe undergoes
several phase transitions. The breaking of the $E_8$ group leads
to the Graviweak Unification (GWU). We have shown that the GWU
action is a special case of the $f(R)$-gravity, and contains not
a Higgs doublet boson of the SM, but a triplet Higgs boson
$\Phi^a$, which leads to the construction of the
black-hole-hedgehog solutions.

\item We have obtained a solution for a black-hole in the region
which contains a global monopole in the framework of the GWU
$f(R)$-gravity. The gravitational field, isovector scalar $\Phi^a$
with $a = 1, 2, 3$, produced by a spherically symmetric
configuration in the scalar field theory, is pointing radially:
$\Phi^a$ is parallel to $\hat{r}$ -- the unit vector in the radial
direction. And in the GWU approach, we obtained a ``hedgehog"
solution (in Alexander Polyakov's terminology). We also showed
that this black-hole-hedgehog solution corresponds to a global
monopole that has been ``swallowed" by a black-hole.

\item We have shown, that hedgehogs having magnetic repulsive
forces of interaction lead to the inflation of the universal
bubble. Such a bubble has a vacuum with a Planck scale VEV:
$v_2\sim 10^{18}$ GeV. This vacuum decays very quickly, and by
this reason is called ``false vacuum".

\item We described a cooling Universe which had a new phase
transition, transforming the ``false" vacuum to the ``true" vacuum.
Hedgehogs confined, and the universal bubble is transformed into
the bubble having spacetime with FLRW-metric and new vacuum
topological defects of $U(1)_{(el\mhyphen mag)}$ group: magnetic vortices
(``strings") by Abrikosov-Nielsen-Olesen, and Sidharth's Compton
wave objects (pointlike defects). These topological defects are
responsible for an almost zero cosmological constant.

\item We have emphasized that, due to the energy conservation law,
the vacuum energy density before the phase transition (for $T >
T_c$) is equal to the vacuum energy density after the phase
transition (for $T < T_c$). Since the vacuum energy densities (DE)
of both vacua are equal, then there is a link between gravitation
and electromagnetism via the Dark Energy.

\item Since $\rho_{vac}$ (at the Planck scale) is almost zero,
then $\rho_{vac}$ (at EW scale) also is almost zero, and we
confirmed a triumph of the Multiple Point Principle predicted the
existence of two universal degenerate vacua with almost zero
vacuum energy density.

\item We also suggested a theory, which predicts the existence of
new bound states (NBS) created by the interaction of the SM Higgs
bosons with 6 top and 6 antitop quarks. Such $6t + 6\bar t$
resonances can be observed by LHC at energy $\sim 1$ TeV.

\item We have investigated whether the EW-vacuum is stable,
unstable, or metastable. It was shown, that hedgehogs, as bound
states, can transform metastability of the EW-vacuum to its
stability. Also we discussed the confirmation of the vacuum
stability by the correction to the Higgs mass coming from the
scalar bound state $S$ of $6t + 6\bar t$.

\item We predicted the production of the $SU(2)$ triplet Higgs
bosons at energy $\sim 10$ TeV which can be detected at LHC.

\item At the end of the paper, we discussed a question: What comes
beyond the Standard Model? Our new cosmological model opens great
possibilities and is predictable: it predicts the mass of
neutrino, solves the hierarchy problem in the SM, predicts that
supersymmetry cannot be detected at LHC, predicts that new
particles - triplet Higgs bosons - can appear at LHC at energy
$\sim$ 10 TeV, predicts that bound states $6t + 6\bar t$ can be
detected at LHC at energies $\sim$ 1 TeV.

\end{enumerate}

\section*{Acknowledgments}
LVL greatly thanks to the B.M. Birla Science Centre (Hyderabad,
India) and personally Prof. B.G. Sidharth, for hospitality,
collaboration and financial support. HBN wishes to thank the Niels
Bohr Institute for the status of professor emeritus and
corresponding support. CRD is thankful to Prof. D.I. Kazakov for
support.

\newpage

\begin{figure}
\includegraphics[height=115mm,keepaspectratio=true,angle=0]{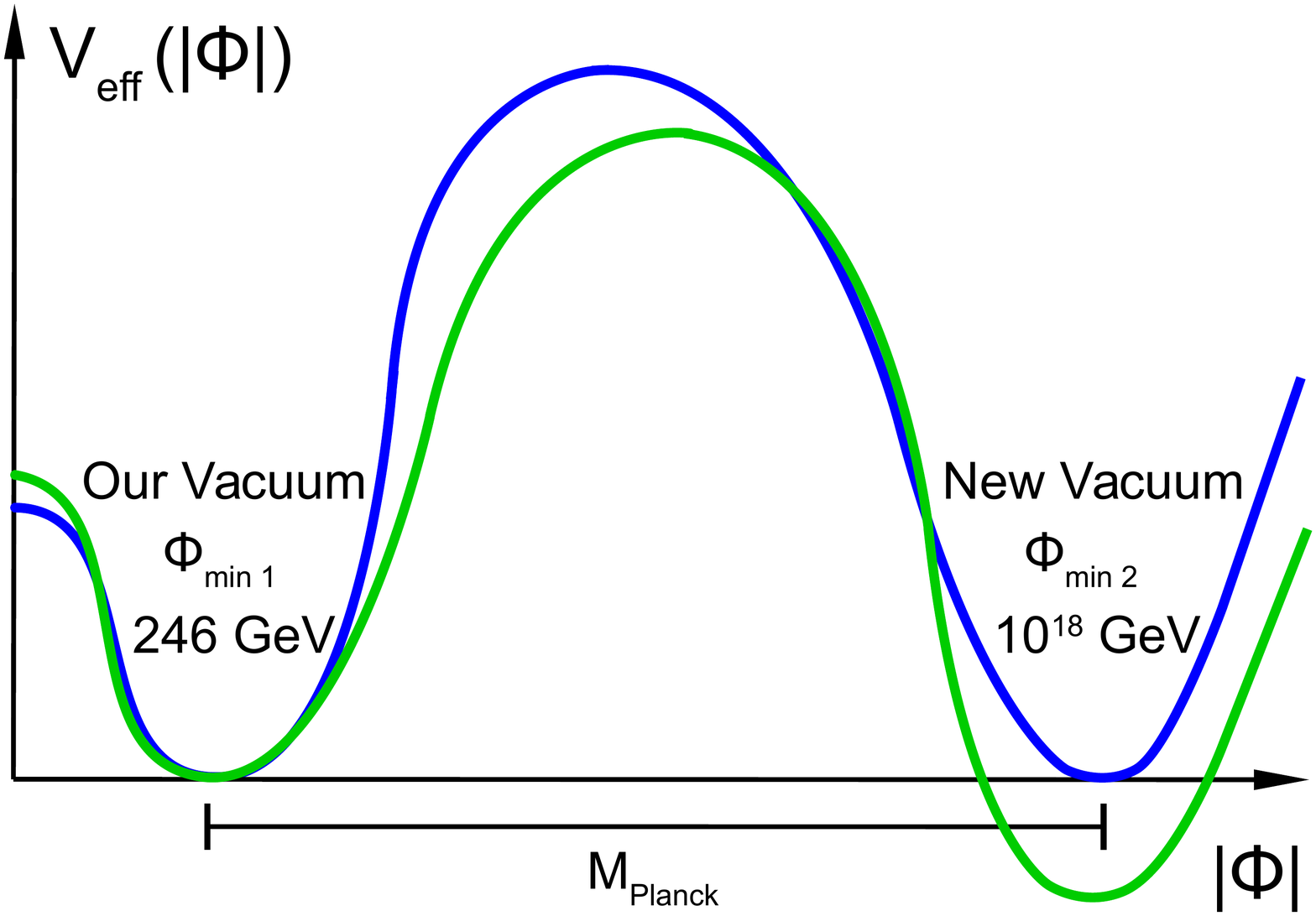}
\caption{It is shown the existence of the second (non-standard)
minimum of the effective Higgs potential in the pure SM at the
Planck scale.} 
\end{figure}

\newpage

\begin{figure}
\includegraphics[height=160mm,keepaspectratio=true,angle=0]{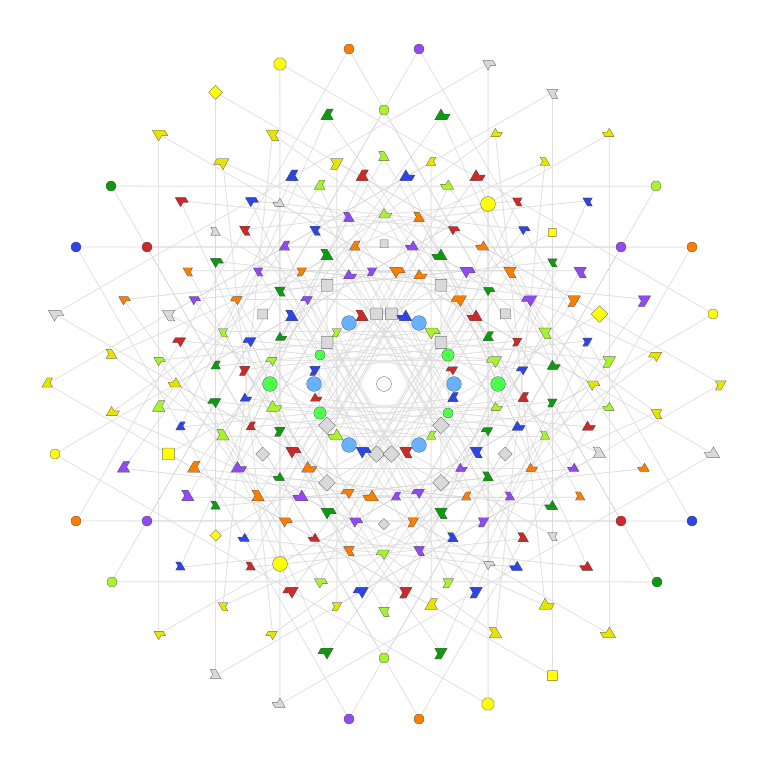}
\caption{Elementary particle states assigned to $E_8$ roots
corresponding to their spin, electroweak, and strong charges
according to $E_8$ Theory.} 
\end{figure}

\newpage

\begin{figure}
\includegraphics[height=105mm,keepaspectratio=true,angle=0]{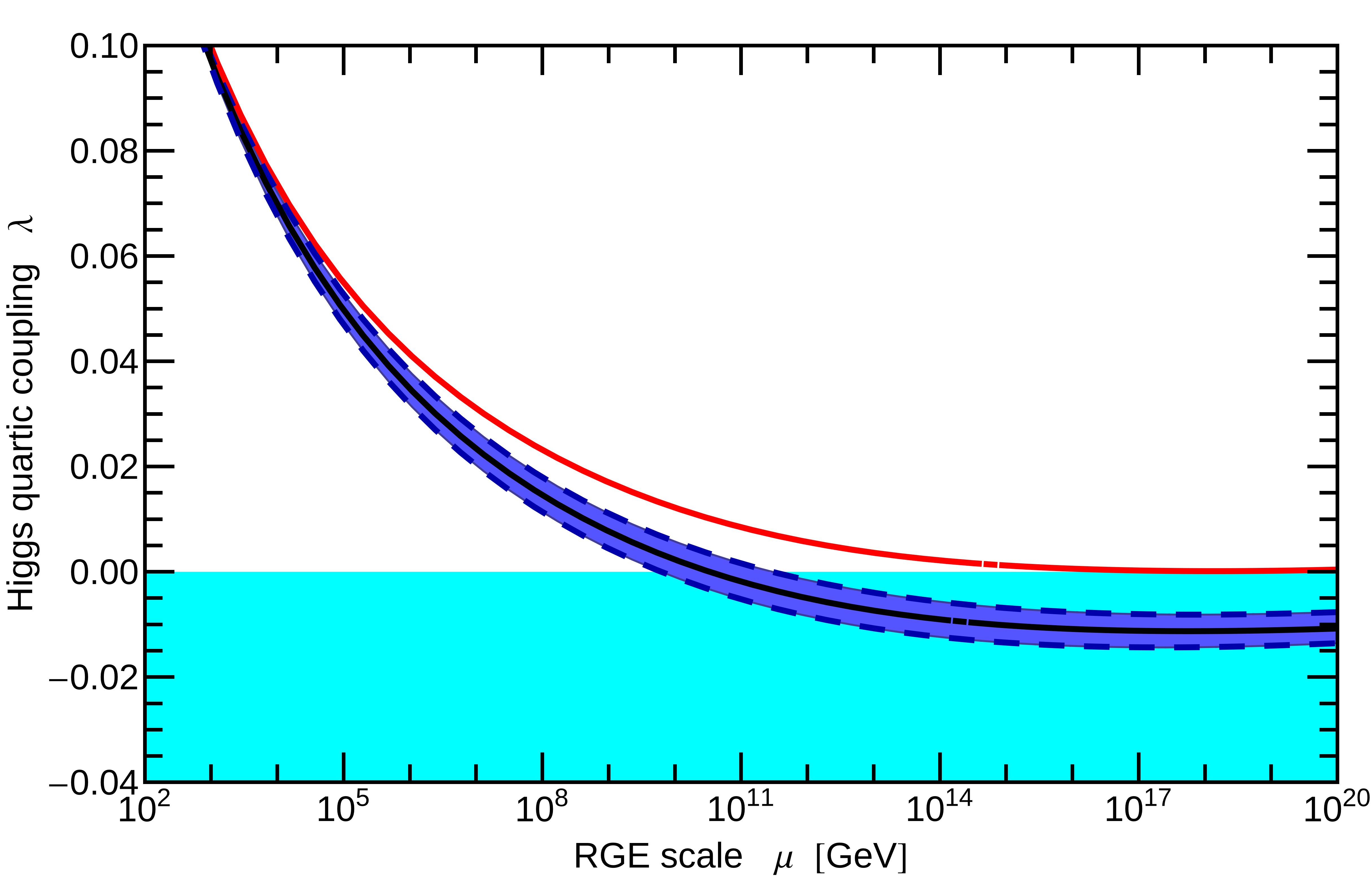}
\caption{RG evolution of $\lambda$.} 
\end{figure}

\newpage

\begin{figure}
\includegraphics[height=85mm,keepaspectratio=true,angle=0]{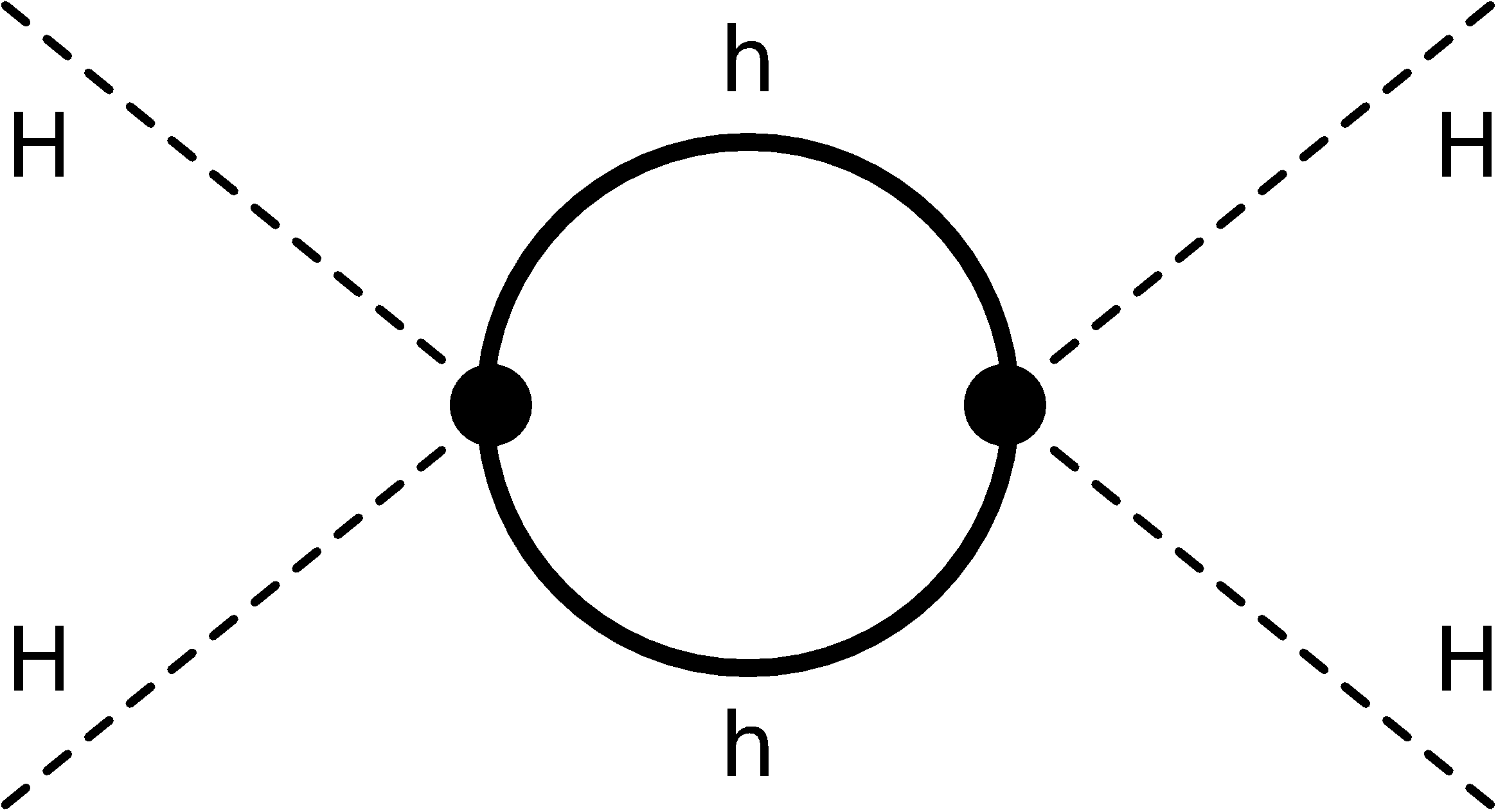}
\caption{Hedgehog $h$ in the loop.} 
\end{figure}

\end{document}